\documentclass{epsconf}
\usepackage{graphicx}
\usepackage{wrapfig}
\usepackage{amsmath}

\title{Effects of the Second Harmonic and Plasma Shaping on the Geodesic Acoustic Mode}
\author{\underline{Johan Anderson}$^1$, Hans Nordman$^1$, Raghvendra Singh$^2$}
\institute{$^1$ Chalmers University of Technology, SE-412 96 G\"{o}teborg, Sweden \\
$^2$ WCI, Center for Fusion Theory, NFRI, Korea and Institute for Plasma Research, Bhat Gandhinagar 2382 428, India
}
\begin{document}
\maketitle
\begin{abstract}
The effects of second harmonics of the density and temperature perturbations  on the linear Geodesic Acoustic Mode (GAM) frequency and non-linear generation of the GAM are investigated, using a fluid model. We show that the second harmonics contribute to the frequency through the density gradient scale length and the wave number of the GAM. In addition, the linear frequency of the GAM is generally increased by coupling to the higher harmonic.
\end{abstract}


\section{Introduction}
Recent experimental work have suggested that the Geodesic Acoustic Mode (GAM) is related to the L-H transition and transport barriers. GAMs are axisymmetric poloidal  $\vec{E} \times \vec{B}$ flows with finite frequencies ($\propto c_s/R$) coupled to up-down-antisymmetric pressure perturbations, which provides a restoring force of the oscillation. The experimental results indicate a periodic modulation of flow and turbulence level with the characteristic limit cycle oscillation at the GAM frequency. In simulations, it was observed that GAMs are only somewhat less effective than the residual zonal flow in providing the non-linear saturation. In particular, it was found in Ref~\cite{sugama2006}, that the damping rates of the GAM are significantly influenced by coupling to higher $m$-modes. The ion temperature gradient (ITG) mode driven by a combination of ion temperature gradients and field line curvature effects is a likely candidate for driving ion scale turbulence, transport and flows. The generation of large scale modes such as zonal flows and GAMs by ITG modes is here realized through the Wave Kinetic Equation (WKE) analysis. This analysis is based on the coupling of the micro-scale turbulence with the GAM through the WKE under the assumptions that there is a large separation of scales in space and time~\cite{chak2007}. In the present work the effects of the second harmonics of the density and temperature perturbations on the linear GAM frequency and non-linear generation of the GAM, using a fluid model~\cite{anderson2002, anderson2013} are investigated. 

We show that the second harmonics contribute to the frequency with higher order terms in $\epsilon_n$ similar to what was discovered in Ref~\cite{elfimov2013}. In addition, effects of plasma shaping will be taken into account~\cite{hager2009}.

\section{The Ion Temperature Gradient Mode}
The description used for toroidal ITG driven modes consists of the ion continuity, ion temperature equations and ion parallel momentum. For simplicity, effects of electron trapping and finite beta effects are neglected in this work. The ion-temperature and ion-continuity equations can be written~\cite{anderson2002}
\begin{eqnarray}
\frac{\partial \tilde{n}}{\partial t} - \left(\frac{\partial}{\partial t} - \alpha_i \frac{\partial}{\partial y}\right)\nabla^2_{\perp} \tilde{\phi} + \frac{\partial \tilde{\phi}}{\partial y} - \epsilon_n g \frac{\partial}{\partial y} \left(\tilde{\phi} + \tau \left(\tilde{n} + \tilde{T}_i \right) \right) + \frac{\partial \tilde{v}_{|| i}}{\partial z}= \nonumber \\
- \left[\phi,n \right] + \left[\phi, \nabla^2_{\perp} \phi \right] + \tau \left[\phi, \nabla^2_{\perp} \left( n + T_i\right) \right] \label{eq:1.03}\\
\frac{\partial \tilde{T}_i}{\partial t} - \frac{5}{3} \tau \epsilon_n g \frac{\partial \tilde{T}_i}{\partial y} + \left( \eta_i - \frac{2}{3}\right)\frac{\partial \tilde{\phi}}{\partial y} - \frac{2}{3} \frac{\partial \tilde{n}}{\partial t} = \nonumber \\
- \left[\phi,T_i \right] + \frac{2}{3} \left[\phi,n \right]. \label{eq:1.04}
\end{eqnarray}
In addition we have used the linear response for the parallel ion motion $\tilde{v}_{|| i} = -i c_s^2/\omega \nabla_{||} (\tilde{\phi} + \tilde{p_i}/\tau)$. With the additional definitions $\tilde{n} = \delta n / n_0$, $\tilde{\phi} = e \delta \phi /T_e$, $\tilde{T}_i = \delta T_i / T_{i0}$ as the normalized ion particle density, the electrostatic potential and the ion temperature, respectively. In the forthcoming equations $\tau = T_i/T_e$, $\omega_{\star} = k_y \vec{v}_{\star} = k_y \rho_s c_s \vec{y}/L_n $, $\rho_s = c_s/\omega_{ci}$ where $c_s=\sqrt{T_e/m_i}$, $\omega_{ci} = eB/m_i c$. We also define $L_f = - \left( d ln f / dr\right)^{-1}$, $\eta_i = L_n / L_{T_i}$, $\epsilon_n = 2 L_n / R$ where $R$ is the major radius and $\alpha_i = \tau \left( 1 + \eta_i\right)$. The perturbed variables are normalized with the additional definitions $\tilde{n} = L_n/\rho_s \delta n / n_0$, $\tilde{\phi} = L_n/\rho_s e \delta \phi /T_e$, $\tilde{T}_i = L_n/\rho_s \delta T_i / T_{i0}$ as the normalized ion particle density, the electrostatic potential and the ion temperature, respectively. The perpendicular length scale and time are normalized to $\rho_s$ and $L_n/c_s$, respectively. Here $\left[ A ,B \right] = \partial A/\partial x \partial B/\partial y - \partial A/\partial y \partial B/\partial x$ is the Poisson bracket. The geometrical quantities will first be considered in the strong ballooning limit ($\theta = 0 $, $g\left(\theta = 0, \kappa \right) = 1/\kappa$~\cite{anderson2002} where $g\left( \theta \right)$ is defined by $\omega_D \left( \theta \right) = \omega_{\star} \epsilon_n g\left(\theta \right)$). Moreover we will also employ semi-local approximation as follows~\cite{anderson2013}, $k_{\parallel} = \left< k_{\parallel} \right>$, $k_{\perp} = \left< k_{\perp} \right>$ and $\omega_D = \left< \omega_D \right>$ will be replaced by the averages defined through the integrals $\left< f(\theta, \kappa, s, q_0) \right> = \frac{1}{N\left(\Psi\right)}\int_{-\pi}^{\pi} d \theta \Psi f(\theta, \kappa, s, q_0) \Psi$ for a simple approximate eigen-function $\Psi(\theta) = \frac{1}{\sqrt{3 \pi}}(1 + \cos \theta)$
\section{The Geodesic Acoustic Mode}
The derivation of the dispersion relation for the Geodesic Acoustic Modes including the $m=2$ higher harmonic coupling to the $m=1$ and $m=0$ components is described here. The GA mode is defined as having $m=n=0$, $q_x \neq 0$ perturbation of the potential field and the $n=0$, $m=1$, $q_x \neq 0$ perturbation in the density, temperatures and the parallel velocity perturbations. In addition we will now consider the $m=2$ components of the density, temperature and parallel velocity perturbations. In total seven coupled differential equations are solved by Fourier components. The GA mode is described by ($q, \omega_q$) and its growth rate is induced by an ITG mode ($k,\omega$). However, only the final GAM dispersion relation will be displayed here,
\begin{equation}
1 - \frac{\epsilon_n^2 q_x^2 \tau}{\omega_q^2 q_{\perp}^2}\left(1 + \frac{2 A}{3 B} \right) \frac{1}{2 C_0} = \frac{q_x^4}{q_{\perp}^2} k_{y}^2 |\tilde{\phi}_k|^2.
\end{equation}
Here $|\tilde{\phi}_k \frac{L_n}{\rho_i}|$ is the saturation level of the ITG turbulence and $A$, $B$ and $C_0$ are coefficients expressing the effects of the higher harmonics. The Wave Kinetic Equation (WKE) analysis is based on the coupling of the micro-scale turbulence with the GAM through the WKE under the assumptions that there is a large separation of scales in space and time~\cite{anderson2002, chak2007}. In this simplified analysis only the Reynolds stress term is retained. Next, we present a short numerical parameter study.
\begin{figure}[ht]
\begin{minipage}[b]{0.5\linewidth}
\centering
\includegraphics[width=6cm, height=4.5cm]{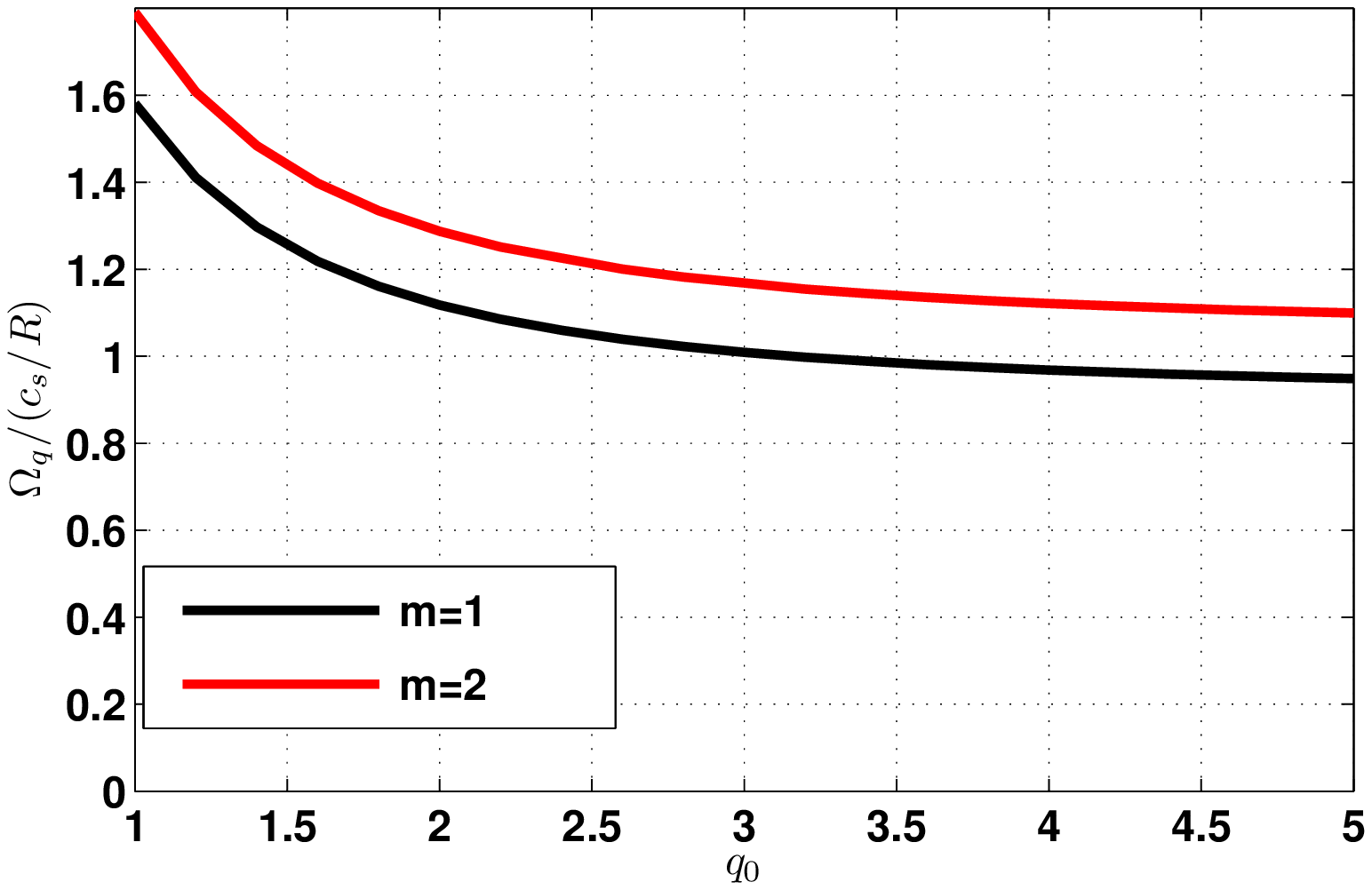}
\caption{The GAM frequency as a function of the safety factor $q_0$ without $m=2$ (black line) and with $m=2$(red line) contributions.}
\label{fig:figure1}
\end{minipage}
\hspace{0.5cm}
\begin{minipage}[b]{0.5\linewidth}
\centering
\includegraphics[width=6cm, height=4.5cm]{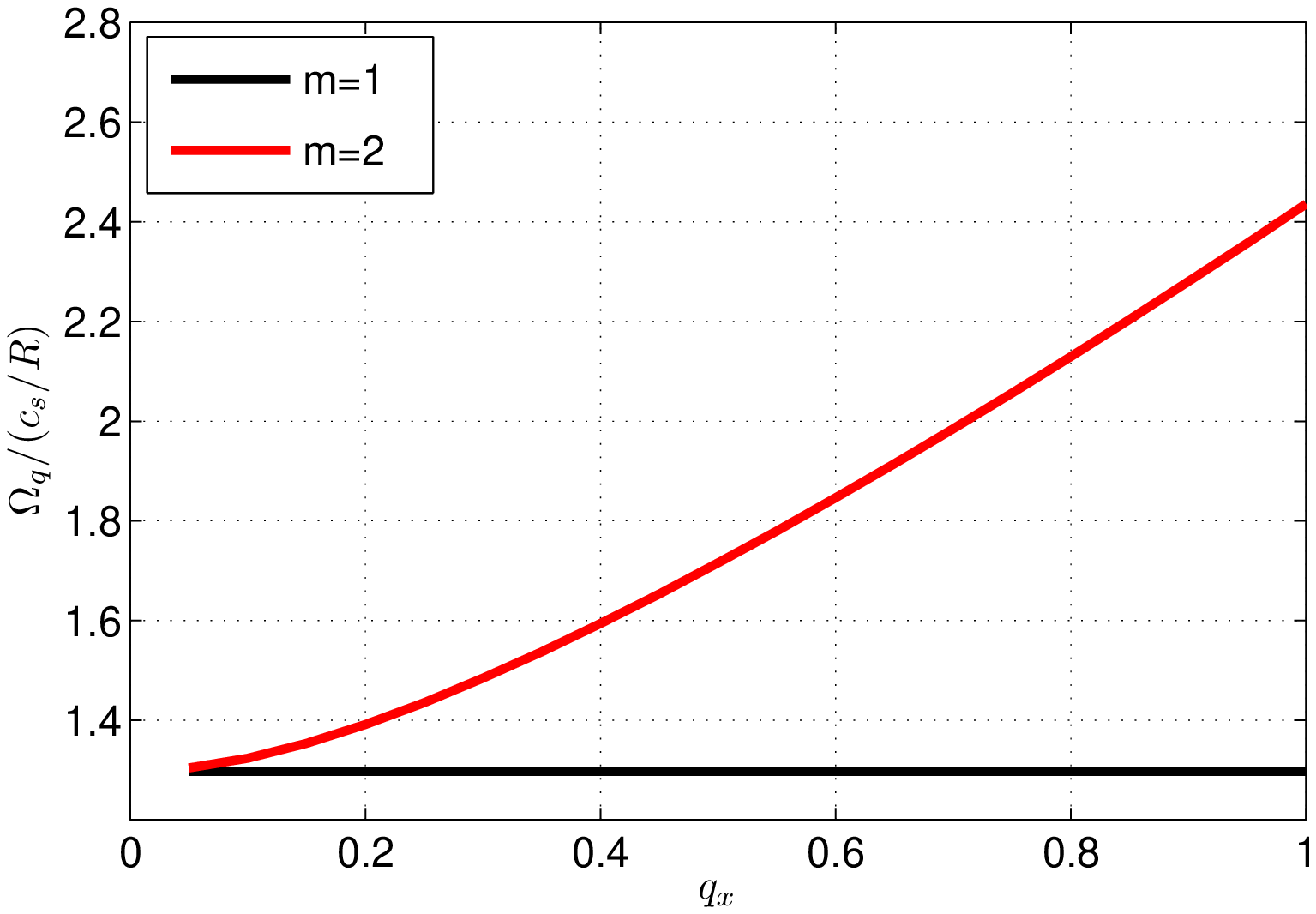}
\caption{The GAM frequency as a function of the wave number $q_x$ without $m=2$ (black line) and with $m=2$(red line) contributions.}
\label{fig:figure2}
\end{minipage}
\end{figure}

\begin{figure}[ht]
\centering
\includegraphics[width=6cm, height=4.5cm]{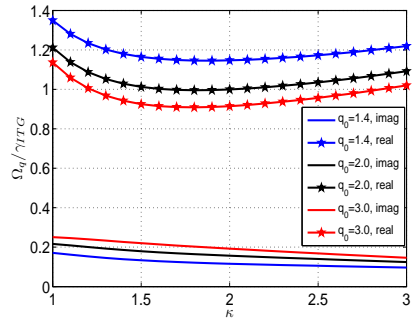}
\caption{Figure 3. The GAM growth rate and frequency as a function of elongation $\kappa$ for safety factors $q_0 = 1.4$ (blue line), $q_0 = 2.0$ (black line) and $q_0 = 3.0$ (red line) with $m=2$ contributions}
\label{fig:figure3}
\end{figure}
In Figure 1, the linear (omitting the right hand side of Eq. (3)) GAM frequency normalized to $c_s/R$ as a function of the magnetic safety factor $q_0$ with contributions from the $m=2$ (red line) mode and without $m=2$ (black line) are shown. The frequency decreases with increasing safety factor $q_0$ in the same manner as found in Ref.~\cite{anderson2013}. The scaling is however independent of $\epsilon_n = 2L_n/R$ and the frequency is increased by about 10\% by including the $m=2$ mode in the analysis. Figure 2 displays the linear GAM frequency normalized by $c_s/R$ as a function of the GAM wave number $q_x$. Note that in the $m=1$ case the frequency is independent of $q_x$ whereas for $m=2$ the frequency sharply increases with increasing $q_x$. In Figure 3, the non-linear contribution on the right hand side of Eq. (3) is included showing the GAM frequency (solid lines with stars) and growth rate (solid lines) normalized to the ITG mode growth rate as a function of the elongation $\kappa$. The effects of geometry is determined by a semi-local analysis similar to the work done in Ref.~\cite{anderson2013} however here modified to allow for varying elongation as well as magnetic shear and safety factor. The other parameters are close to those of the cyclone base case $\eta_i = 3.14$, $\epsilon_n = 0.909$, $s = 0.8$ and $k_{\perp}^2 = 0.1$. The effects of elongation is weakly stabilizing the GAM by reducing the growth rate whereas the frequency first decreases and then increases for larger values of $\kappa$. Here the mode coupling saturation level is assumed ~\cite{anderson2002}.
\section{Summary}
A system of equations is derived which describes the dynamics of the Geodesic Acoustic Mode (GAM) including the coupling to higher harmonics ($m=2$). The Wave-Kinetic (WK) formalism is used to derive the driving term for the GAM growth rate. It has been suggested in experimental work that the GAM is related to the L-H transition and transport barriers making it pertinent to analytically explore the properties of the GAM in detail. In addition, we have included the effects of plasma shaping through a semi-local analysis, however due to the complexity of the resulting integrals closed analytical forms are not available. We have found that the GAM frequency is significantly altered by including the coupling to higher harmonics, generally increasing the frequency. The effects of elongation is less decisive however a weakly stabilizing effect is present. In future work work the sub-dominant non-linear interactions will be investigated as well as more parameters studies.

\end{document}